\documentclass[aps,prb,twocolumn,english,showpacs,superscriptaddress,amssymb,amsfonts]{revtex4}
\usepackage{amsmath}
\usepackage{dsfont}
\usepackage{amstext}

\usepackage{amssymb}
\usepackage{amsbsy}
\usepackage{amsthm}
\usepackage{epsfig}
\usepackage{graphicx}
\usepackage{bbm}
\usepackage{hyperref}
\usepackage{color}

\newcommand{\Ref}[1]{Ref.~\onlinecite{#1}}

\newcommand{\bst}{{\boldsymbol{T}}}
\newcommand{\bse}{{\boldsymbol{e}}}
\newcommand{\bsg}{{\boldsymbol{g}}}

\newcommand{\ie}{{\emph{i.e.~}}}
\makeatletter

\newcommand{\Rmnum}[1]{\expandafter\@slowromancap\romannumeral #1@}
\makeatother
\newcommand{\imth}{\hspace{1pt}\mathrm{i}\hspace{1pt}}

\newcommand{\eg}{{\emph{e.g.~}}}

\newcommand{\mbz}{{\mathbb{Z}}}

\newcommand{\bea}{\begin{eqnarray}}
\newcommand{\eea}{\end{eqnarray}}
\newcommand{\bpm}{\begin{pmatrix}}
\newcommand{\epm}{\end{pmatrix}}
\newcommand{\bal}{\begin{aligned}}
\newcommand{\eal}{\end{aligned}}
\def\beq{\begin{equation}}
\def\eeq{\end{equation}}
\newcommand{\tp}{\tilde\phi}

\begin{document}
\title{Gapped symmetric edges of symmetry protected topological phases}

\author{Yuan-Ming Lu}
\affiliation{Department of Physics, University of California, Berkeley, CA 94720, USA}
\affiliation{Materials Science Division, Lawrence Berkeley National Laboratories, Berkeley, CA 94720}

\author{Dung-Hai Lee}
\affiliation{Department of Physics, University of California, Berkeley, CA 94720, USA}
\affiliation{Materials Science Division, Lawrence Berkeley National Laboratories, Berkeley, CA 94720}

\begin{abstract}
Symmetry protected topological (SPT) phases are gapped quantum phases which host symmetry-protected gapless edge excitations. On the other hand, the edge states can be gapped by spontaneously breaking symmetry. We show that topological defects on the symmetry-broken edge cannot proliferate due to their fractional statistics. A gapped symmetric boundary, however, can be achieved between an SPT phase and certain fractionalized phases by condensing the bound state of a topological defect and an anyon. We demonstrate this by two examples in two dimensions: an exactly solvable model for the boundary between topological Ising paramagnet and double semion model, and a fermionic example about the quantum spin Hall edge. Such a hybrid structure containing both SPT phase and fractionalized phase generally support ground state degeneracy on torus.
\end{abstract}

\maketitle

\section{Introduction}

Topological insulators\cite{Hasan2010,Moore2010,Qi2011} (TIs) support gapless boundary excitations in spite of a gapped bulk spectrum. The edge states are believed to be stable against any perturbation, as long as certain symmetries are preserved. When symmetries are broken, however, TIs can be continuously tuned into a trivial atomic insulator without phase transitions. Recently it's realized\cite{Chen2013a} that aside from weakly-interacting electrons, such phases generally exist in interacting bosons and they are dubbed ``symmetry protected topological'' (SPT) phases.

When symmetries are spontaneously broken, a gap can open up in the edge spectrum of SPT phases. There are always topological defects (kinks)\cite{Mermin1979} associated with spontaneous symmetry breaking, such as the domain wall excitation in an Ising ferromagnet. Usually by proliferating the defects one can restore symmetry, leading to a gapped symmetric state: \eg the disordered phase of a transverse Ising model can be achieved by ``condensing'' the domain walls. Similarly can one achieve a gapped symmetric state on the edge of an SPT phase?

In this work we answer this question constructively, focusing on two spatial dimensions (2+1-D). We show that topological defects (kinks) on the boundary always carry fractional statistics\cite{Wilczek1990B} or symmetry quantum numbers, hence their proliferation is either forbidden or breaks symmetry. However, on a boundary between SPT phase and certain fractionalized phase (which hosts anyon excitations\cite{Wilczek1990B}), one can form a bosonic bound state of the kink on SPT side and anyon on fractionalized side. Proliferating this composite object will lead to a gapped symmetric boundary. This can be generalized to any spatial dimensions. Two examples are presented: 1) boundary between bosonic $Z_2$-SPT and double semion model, equipped with an exactly solvable model; 2) boundary between quantum spin Hall insulator (QSHI) and a fractionalized QSHI$^\ast$ phase. We show that a hybrid structure containing SPT and fractionalized phases support ground state degeneracy (GSD) on a torus (FIG. \ref{fig:torus}).

\section{Edge field theory and ``fractional'' defects of SPT phases}

 SPT phases\cite{Chen2013a} in two spatial dimensions can be described\cite{Lu2012a} by multi-component Chern-Simons theory\cite{Read1990,Wen1992,Frohlich1991} with a symmetric unimodular matrix ${\bf K}$. In particular, 2+1-D SPT phases host gapless edge excitations, described by chiral bosons $\{\phi_i\}$ with the following effective field theory\cite{Wen1995}:
\bea\label{edge action}
\mathcal{L}_{edge}=\sum_{I,J}\frac1{4\pi}{\bf K}_{I,J}\partial_t\phi_I\partial_x\phi_J-{\bf V}_{I,J}\partial_x\phi_I\partial_x\phi_J
\eea
where ${\bf V}$ is a positive-definite real symmetric matrix. Backscattering terms $\sim\cos(\sum_I l_I\phi_I)$ are generally not allowed by symmetry\cite{Lu2012a,Levin2012a} denoted by group $G_s$.

A simple example is topological paramagnet protected by $Z_n$ symmetry\cite{Levin2012,Lu2012a} where $G_s=Z_n\equiv\{\bsg,\bsg^2,\cdots,\bsg^N=\bse\}$. Its edge structure is characterized by ${\bf K}=\bpm0&1\\1&0\epm$ in effective theory (\ref{edge action}), where under $Z_n$ symmetry operation the chiral bosons transform as\cite{Lu2012a}
\bea\label{sym transf:Zn}
\bpm\phi_1\\ \phi_2\epm\overset{\bsg}\longrightarrow\bpm\phi_1\\ \phi_2\epm+\frac{2\pi}{n}\bpm1\\k\epm,~~~1\leq k\leq n-1.
\eea
Here backscattering terms $\mathcal{H}_{bs}\sim\cos(\phi_{1,2}-\alpha_{1,2})$ are forbidden by the above $Z_n$ symmetry. Once symmetry is broken, edge states in (\ref{edge action}) can be gapped out by $\mathcal{H}_{bs}$.

Spontaneous symmetry breaking on the edge will pin chiral boson fields $\phi_{1,2}$ at certain classical values. Distinct classical values $\langle\phi_{1.2}\rangle$ correspond to different ways to break discrete $Z_n$ symmetry, and there are topological defects $-$ domain walls (or kinks) which spatially separates these different ``vacua''. Different kinks are classified\cite{Mermin1979} by homotopy group $\pi_0(Z_n)=Z_n$, \ie there are $n$ distinct types of domain walls, including the trivial one $-$ no domain wall. All stable kinks can be generated by a fundamental domain wall, which in our case $G_s=Z_n$ can be written in terms of chiral bosons\cite{Chen2014}:
\bea\label{domain wall:Zn}
\hat{D}_{Z_n|k}(x)=\exp\Big[\imth\frac{k\phi_1(x)+\phi_2(x)}{n}\Big]
\eea
As implied by (\ref{edge action}) chiral bosons $\phi_{1,2}$ obey commutation relation
$[\phi_1(x),\phi_2(y)]=2\pi\imth\cdot\theta(y-x)$.
It's straightforward to see the classical values $\langle\phi_{1,2}\rangle$ on two sides of domain wall (\ref{domain wall:Zn}) are related by the symmetry transformation (\ref{sym transf:Zn}), \eg $\hat{D}_{Z_n|k}(x)\phi_1(y)\hat{D}_{Z_n|k}^{-1}(x)=\phi_1(y)+\frac{2\pi}{n}\theta(x-y)$.

A natural question is: can we restore symmetry ($G_s=Z_n$ here) simply by proliferating topological defects (kinks) and obtain a gapped symmetric edge of SPT phases? Remarkably the kink (\ref{domain wall:Zn}) is neither
a boson nor a fermion: generally it obeys fractional statistics\cite{Wilczek1990B,Wen1995}
\bea
\hat{D}_{Z_n|k}(x)\hat{D}_{Z_n|k}(y)=\hat{D}_{Z_n|k}(y)\hat{D}_{Z_n|k}(x)e^{\imth\frac{2\pi k}{n^2}\text{Sgn}(x-y)}\notag
\eea
with statistical angle $\theta_{Z_n|k}=2\pi\frac{k}{n^2}$. Moreover it has fractional mutual statistics with bosonic excitations $\{e^{\imth\phi_{1,2}}\}$:
\bea
&\notag\hat{D}_{Z_n|k}(x)e^{\imth\phi_1(y)}=e^{\imth\phi_1(y)}\hat{D}_{Z_n|k}(x)e^{\imth\frac{2\pi}{n}\theta(x-y)},\\
&\notag\hat{D}_{Z_n|k}(x)e^{\imth\phi_2(y)}=e^{\imth\phi_2(y)}\hat{D}_{Z_n|k}(x)e^{-\imth\frac{2\pi k}{n}\theta(y-x)}.
\eea
Its fractional statistics leads to destructive interference in the path integral when domain walls proliferate, therefore suppressing instanton events which create/annihilate domain walls. As a result in $Z_n$-SPT phases, it is impossible to disorder the symmetry-broken edges simply by proliferating solitons. More generally on the boundaries of $d+1$-D SPT phases, there are similar obstructions to proliferate symmetry-breaking topological defects\cite{Vishwanath2013}.

\section{Gapped symmetric boundary between a SPT phase and a fractionalized phase}

On the other hand, on the boundary between a SPT phase $\mathcal{S}$ and a fractionalized phase $\mathcal{F}$ (\ie intrinsic topological order\cite{Wen2004B} which supports anyon excitations in the bulk), the bound state of a kink from SPT side and an anyon from fractionalized side can have bosonic statistics. This object in principle can proliferate and gap out all boundary excitations, restoring symmetry $G_s$. Notice that fractionalized phase $\mathcal{F}$ should also respect symmetry $G_s$, in other words it is a symmetry enriched topological (SET) phase\cite{Essin2013,Mesaros2013,Hung2013,Lu2013,Hung2013a}.

Given a 2+1-D SPT phase $\mathcal{S}$, only certain 2+1-D fractionalized phases can have a gapped symmetric boundary with $\mathcal{S}$. For example edge states of $\mathcal{F}$ and $\mathcal{S}$ must have the same number of chiral edge modes, \ie same chiral central charge $c_-$. Meanwhile $\mathcal{F}$ must support anyon excitations with the same statistics and symmetry quantum numbers as topological defects (kinks) on the edge of $\mathcal{S}$.
We propose the following conjecture, which provides a way to look for fractionalized phase $\mathcal{F}$ sharing a gapped symmetric boundary with SPT phase $\mathcal{S}$:

\emph{An SPT phase $\mathcal{S}$ in any spatial dimensions always possesses a gapped symmetric boundary with a fractionalized (SET) phase $\mathcal{F}$, where $\mathcal{F}$ is obtained by gauging\cite{Levin2012} an Abelian discrete symmetry in $\mathcal{S}$.}
\bea\label{conjecture}\eea

In the following we give a proof of this conjecture in 2+1-D based on Chern-Simons approach. 

An Abelian discrete symmetry group always has the form of a direct product of cyclic groups:
\bea\label{abelian discrete group}
G_s=\prod_{n\geq2}\big(Z_{n}\big)^{\alpha_n},~~~\alpha_n=0,1,2,\cdots.
\eea
Here we explicitly prove the conjecture for the case of a generic $Z_n$ symmetry (cyclic group of order $n$).

Consider an arbitrary 2+1-D SPT phase $\mathcal{S}$ with $Z_n$ symmetry. Since it's an Abelian 2+1-D phase its bulk effective theory is a multi-component Chern-Simons theory\cite{Wen2004B}
\bea\label{bulk action}
\mathcal{L}_{bulk}=\frac{\epsilon^{\mu\nu\rho}}{4\pi}\sum_{I,J}a_\mu^I{\bf K}_{I,J}\partial_\nu a^J_\rho-\sum_I a_\mu^I j^\mu_I
\eea
where $j^\mu_I$ are the quasiparticle currents. In the long-wavelength limit its edge excitations are described by\cite{Wen1995,Lu2012a}:
\bea\label{edge action}
\mathcal{L}_{edge}=\sum_{I,J}\frac1{4\pi}{\bf K}_{I,J}\partial_t\phi_I\partial_x\phi_J-{\bf V}_{I,J}\partial_x\phi_I\partial_x\phi_J
\eea
where ${\bf K}$ is a unimodular matrix and ${\bf V}$ is a real positive-definite matrix. Denoting the generator of $Z_n$ symmetry by $\bsg$ ($\bsg^n=\bse$), the edge chiral bosons $\{\phi_i\}$ transforms as\cite{Lu2012a,Lu2013}
\bea\label{sym transf}
\phi_I\overset{\bsg}\longrightarrow\phi_I+\delta\phi^\bsg_I;~~~~~n{\bf K}\vec{\delta\phi^\bsg}=0\mod2\pi.
\eea

First of all, what is the fractionalized (SET) phase $\mathcal{F}$ after gauging $Z_n$ symmetry? A well-defined way to gauge the symmetry in a lattice model\cite{Levin2012} is to couple the local degrees of freedom (which lives on lattice sites and transforms under $Z_n$ symmetry) to a dynamical $Z_n$ gauge field (which lives on links). In an effective field theory, the effect of gauging a symmetry is captured by deconfining the symmetry twist (or symmetry flux)\cite{Levin2012,Hung2012a,Lu2013} in the original SPT phase.
A symmetry twist has the following property: when a particle carrying symmetry quantum numbers goes around this symmetry twist once, it transforms under a symmetry operation. In the language of gauge field theory (\ref{bulk action}), particles are integer gauge charges of fields $\{a_\mu^I\}$: they are labeled by an integer vector ${\bf l}$ and represented by operators $e^{\imth\sum_I{\bf l}_I\phi_I}$ on the edge. Notice that particle $e^{\imth\phi_I}$ pick up a phase $e^{\imth\delta\phi^\bsg_I}$ under $Z_n$ symmetry operation $\bsg$, therefore symmetry twists are nothing but gauge fluxes in (\ref{bulk action}). In a Chern-Simons theory, gauge fluxes are also gauge charges, which becomes transparent in the following equations of motion
\bea
\notag\frac{\delta\mathcal{L}_{bulk}}{\delta a_\mu}=0\Longrightarrow j^\mu_I=\frac{\epsilon^{\mu\nu\rho}}{2\pi}\sum_{J}{\bf K}_{I,J}\partial_\nu a^J_\rho
\eea
As a result, in the fractionalized phase $\mathcal{F}$ obtained by gauging $Z_n$ symmetry, quasiparticles corresponding to (fractional gauge charge) vector
\bea
{\bf l}^\bsg={\bf K}\vec{\delta\phi^\bsg}/2\pi,
\eea
are new excitations in SET phase $\mathcal{F}$. One can immediately see from (\ref{sym transf}) that $n{\bf l}^\bsg$ must be an integer vector. Moreover in a $Z_n$-SPT the symmetry transformations (\ref{sym transf}) form a faithful representation\cite{Lu2012a} of $Z_n$ group, meaning that at least one component of integer vector ${\bf l}^\bsg$ is $1/n$. Without loss of generality, we assume that ${\bf l}^\bsg_{p+1}=1/n$ where $\dim{\bf K}=p+1$. In other words we have
\bea
{\bf l}^\bsg=({\bf v}^T,1)^T/n,~~~{\bf v}\in\mbz^p.
\eea
As discussed in \Ref{Lu2013}, the fractionalized SET phase $\mathcal{F}$ obtained by gauging $Z_n$ symmetry is described by Chern-Simons theory (\ref{bulk action}) with matrix
\bea
{\bf K}^\bsg={\bf M}^{-1}{\bf K}\big({\bf M}^{-1}\big)^T,~~~{\bf M}=\bpm1_{p\times p}&{\bf v}/n\\0_{1\times p}&1/n\epm\notag.
\eea
It's easy to see
\bea
{\bf M}^{-1}=\bpm1_{p\times p}&-{\bf v}\\0_{1\times p}&n\epm,~~~\det{\bf K}^\bsg=n^2\det{\bf K}.
\eea
We label chiral bosons on the edge of SET phase $\mathcal{F}$ by $\{\tp_I|1\leq I\leq p+1\}$. They have the following correspondence with the edge chiral bosons $\{\phi_i\}$ in SPT phase $\mathcal{S}$:
\bea
\tp_I\leftrightarrow\sum_{J}{\bf M}_{J,I}\phi_I
\eea
Similar relations hold for their symmetry transformations $\{\tilde\delta\phi^\bsg_I\}$ and $\{\delta\phi^\bsg_I\}$.

In SET phase $\mathcal{F}$, mutual statistics between particle ${\bf l}$ and the new (fractional) particle ${\bf l}^\bsg$ is given by\cite{Wen2004B}:
\bea
\theta_{{\bf l},{\bf l}^\bsg}=2\pi{\bf l}^T{\bf K}^{-1}{\bf l}^\bsg={\bf l}^T\vec{\delta\phi^\bsg}.
\eea
Generally $\delta\phi^\bsg_I/2\pi=p_I/q_I$ where $(p_I,q_I)$ are two mutually prime integers. It's not hard to check that the following set of backscattering terms between the SPT edge and SET edge gives rise to a gapped symmetric boundary between them:
\bea
\notag\mathcal{L}_1=\sum_{I=1}^p\cos\big[q_I(\tp_I-\phi_I)\big]+\cos\big[n(\tp_{p+1}-\sum_I{\bf l}^\bsg_I\phi_I)\big].
\eea\\

In the case of gauging a continuous symmetry, the conjecture (\ref{conjecture}) is not valid anymore. An obvious example is $U(1)$-SPT in 2+1-D, \ie bosonic integer quantum Hall effect\cite{Lu2012a}. The chiral central charge is always $c_-=0$ for such a $U(1)$-SPT phase $\mathcal{S}$. After gauging $U(1)$ symmetry, we obtain a fractionalized phase $\mathcal{F}$ described by $U(1)$ level-$\sigma_{xy}$ Chern-Simons term, which has chiral central charge $c_-=\text{Sgn}(\sigma_{xy})$. Apparently there cannot be a gapped edge between $\mathcal{S}$ and $\mathcal{F}$.

\section{Examples}

In this section we'll demonstrate the conjecture (\ref{conjecture}) in two examples.

\subsection{Topological Ising paramagnet $\Big(G_s=Z_2\Big)$}

The simplest example of 2+1-D bosonic SPT phases is the topological Ising paramagnet\cite{Levin2012} with $G_s=Z_2$ symmetry. In its edge effective theory (\ref{edge action}), ${\bf K}=\bpm0&1\\1&0\epm$ and Ising symmetry acts on chiral bosons $\phi_{1,2}$ as (\ref{sym transf:Zn}) with $n=2,k=1$. The domain wall operator $\hat{D}_{Z_2|1}$ in (\ref{domain wall:Zn}) is a semion\footnote{Strictly speaking the domain wall could also be an anti-simion, as a bound state of a semionic domain wall and a local spin flip.}, with fractional statistics $\theta=\pi/2$. In order to acquire a gapped symmetric edge, the corresponding fractionalized (SET) phase must support bulk semion excitations and non-chiral edge states. The simplest choice is the double semion state\cite{Levin2005} with ${\bf K}=\bpm2&0\\0&-2\epm$ in (\ref{edge action}). Note that gauging Ising symmetry in $Z_2$-SPT leads to\cite{Levin2012} nothing but double semion model. On its edge there are two branches of chiral bosons $\phi_{s,\bar{s}}$, where $e^{\imth\phi_s}$ creates a semion and  $e^{\imth\phi_{\bar{s}}}$ creates an anti-semion. Now the following tunneling terms between the SPT edge ($\phi_{1,2}$) and double semion edge ($\phi_{s,\bar{s}}$)
\bea\label{boundary term:Z2}
\mathcal{H}_{t}=T_1\cos(\phi_1+\phi_2-2\phi_s)+T_2\cos(\phi_1-\phi_2-2\phi_{\bar s}).
\eea
can open up a gap on their boundary without breaking Ising symmetry. Here spins in double semion model doesn't transform under Ising symmetry. In contrast, on the ``pure" boundary between this $Z_2$-SPT phase and the vacuum, there is no way to get rid of gapless excitations without breaking symmetry\cite{Chen2013a,Levin2012}.
\begin{figure}
 \includegraphics[width=0.4\textwidth]{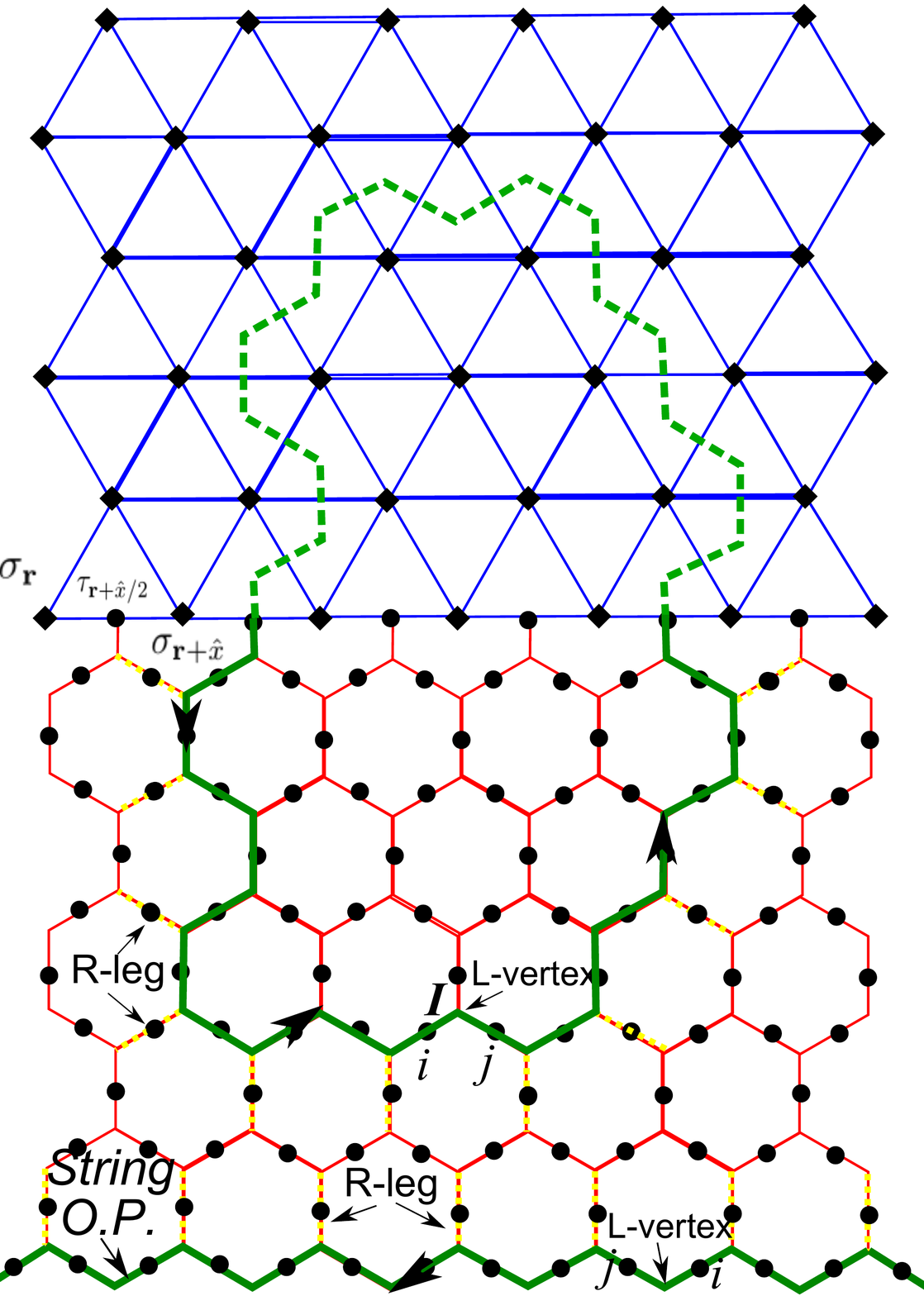}
\caption{(color online) Exactly solvable model of a gapped symmetric edge between bosonic $Z_2$-SPT (diamonds denote spins $\{\vec\sigma_{\bf r}\}$ on triangular lattice) and double semion model (solid circles denote spins $\{\vec\tau_{{\bf r}^\prime}\}$ on edge centers of honeycomb lattice). Dashed green lines on triangular lattice represent domain wall configurations in ground state wavefunction of Ising paramagnet, using $\sigma^z=\pm1$ basis. Solid green lines denote (oriented) string-net configurations in the ground state of double semion model, where $\tau^z=1$ correspond to no string and $\tau^z=-1$ to occupied by a string. The boundary term (\ref{boundary ham:Z2}) guarantees that a $Z_2$ domain wall from upper half always forms a bound state with the end of a string on the lower half. Such a bosonic bound state can hop and condense, giving rise to a gapped symmetric edge.}\label{fig:model}
\end{figure}

In the following we present an exactly solvable model for such a boundary between bosonic $Z_2$-SPT\cite{Levin2012} and double semion model\cite{Levin2005}. Its Hamiltonian consists of commuting local projectors:
\bea\label{model}
 H=H_{{SPT}}\{\sigma_{\bf a}\}+H_{{SET}}\{\tau_{{\bf i}}\}+H_{BDY},
\eea
On $Z_2$-SPT side\cite{Levin2012}
\bea
\notag H_{{SPT}}=\sum_{\bf a}\sigma_{\bf a}^x\big(\prod_{\langle{\bf a,b,c}\rangle}\imth^{\frac{1-\sigma_{\bf b}^z\sigma^z_{\bf c}}{2}}\big)
\eea
where $\langle{\bf a,b,c}\rangle$ runs over all six (nearest neighbor) triangles containing ${\bf a}$. On the side of double semion model\cite{Levin2005}
\bea
\notag H_{SET}=-\sum_{{\bf I}}\prod_{\text{legs of}~{\bf I}}\tau_{\bf i}^x+\sum_{\bf r}(\prod_{\text{edges of}~{\bf r}}\tau^z_{\bf j})(\prod_{\text{R-legs of}~{\bf r}}\imth^{\frac{1-\tau^x_{\bf j}}{2}})
\eea
{\bf I} denotes vertices and ${\bf r}$ denotes hexagonal plaquette center. The boundary Hamiltonian connects the two sides symmetrically
\begin{align}\label{boundary ham:Z2}
& H_{BDY}=\sum_{\bf r\in \text{boundary}}-\tau^x_{{\bf r}+\hat{x}/2}\sigma^z_{\bf {r}}\sigma^z_{{\bf r}+\hat{x}}\\
&\notag+\sigma^x_{\bf r}(\prod_{\langle{\bf r,b,c}\rangle}\imth^{\frac{1-\sigma_{\bf b}^z\sigma^z_{\bf c}}{2}})(\prod_{\text{edges of}~{\bf r}}\tau^z_{\bf j})(\prod_{\text{R-legs of}~{\bf r}}\imth^{\frac{1-\tau^x_{\bf j}}{2}})
\end{align}
as shown in FIG. \ref{fig:model}. It's easy to verify that all terms commute with each other. Physically the 1st term in $H_{BDY}$ guarantees a domain wall from $Z_2$-SPT side is always bound to a string from double semion side, while the 2nd term provides kinetic energy which allows this bound state to hop on the boundary. Therefore they can ``condense" on the boundary and gap out the edge states without breaking Ising symmetry. The low-energy effective theory for boundary Hamiltonian $H_{BDY}$ is nothing but (\ref{boundary term:Z2}).

\begin{figure}
 \includegraphics[width=0.4\textwidth]{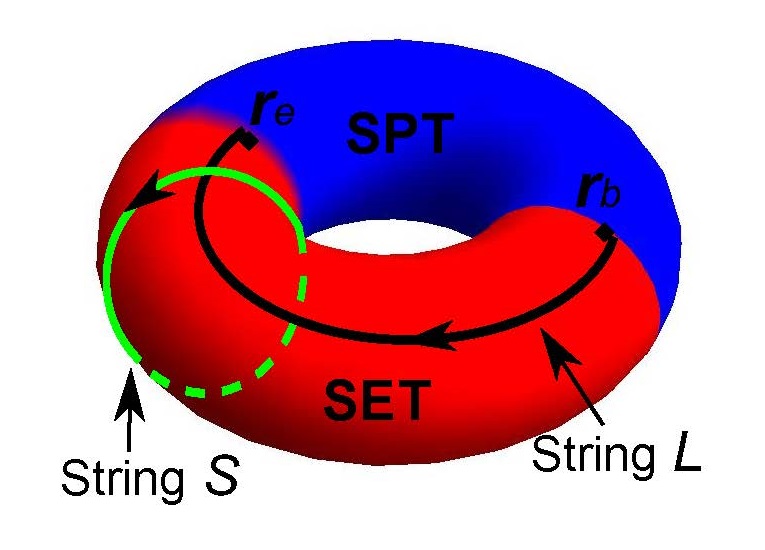}
\caption{(color online) Illustration of the hybrid structure on a torus, containing SPT phase $\mathcal{S}$ (blue) and fractionalized SET phase $\mathcal{F}$ (red). Both boundaries between the two regions are fully gapped without breaking any symmetry. The two oriented loops correspond to string order parameters $\hat{O}_S$ (green) and $\hat{O}_L$ (black). Unlike $\hat{O}_S$ which consists of operators only from the SET region, open string operator $\hat{O}_L$ also contains operators from SPT side: such as spin $\sigma^z$ on sites ${\bf r}_b$ and ${\bf r}_e$ in the example of topological Ising paramagnet. Both string operators preserve the symmetry of the system.}\label{fig:torus}
\end{figure}

The Ising symmetry in this model is implemented by $\bsg=\prod\sigma^x$, thus $\vec\tau$ spins in double semion model is invariant under Ising spin flip. Now let's consider model (\ref{model}) on a torus (see FIG. \ref{fig:torus}), where a half of the torus hosts double semion model and the other half hosts $Z_2$-SPT phase. Both shared boundaries between $Z_2$-SPT and double semion model are gapped by Hamiltonian (\ref{boundary ham:Z2}). However there is a 2-fold ground state degeneracy (GSD) in such a hybrid structure on torus. This can be easily verified by comparing the number of independent stablizers\cite{Kitaev2003} (local commuting projectors in $H$) and the total number of spins, since there is a global constraint for the local projectors in model (\ref{model}) on torus:
\bea
\notag\prod_{{\bf I}}(\prod_{\text{legs of}~{\bf I}}\tau_{\bf i}^x)\cdot\prod_{\bf r\in \text{boundary}}(\tau^x_{{\bf r}+\hat{x}/2}\sigma^z_{\bf {r}}\sigma^z_{{\bf r}+\hat{x}})=1
\eea
These two degenerate ground states can be labeled by eigenvalues of string ``order parameter''\cite{Levin2005}
\bea
\notag\hat{O}_S=\prod_{\text{edges of}~S}\tau^z_{\bf j}\prod_{\text{R-legs}}\imth^{\frac{1-\tau_{\bf j}^x}{2}}\prod_{\text{L-vertices}}(-1)^{\frac{(1-\tau_{\bf i}^x)(1+\tau_{\bf j}^x)}{4}}.
\eea
The closed string $S$ winds around a non-contractible loop of torus once, parallel to the boundary between $Z_2$-SPT and double semion model, as illustrated by the horizontal green loop in FIG. \ref{fig:model}. It's easy to verify $(\hat{O}_S)^2=1$, hence it has eigenvalues $\pm1$.

A pure double-semion model on torus has 4-fold GSD labeled by two commuting string operators: $\hat{O}_S$ and $\hat{O}^\prime_S=\prod_{\text{R-legs}}\tau_{\bf j}^x$. However in the ground states of our model, the 1st term of boundary Hamiltonian (\ref{boundary ham:Z2}) fixes the eigenvalue of $\hat{O}^\prime_S$ to be 1. Therefore the original 4-fold GSD of double-semion model on torus reduces to 2-fold in the hybrid structure here. The two degenerate ground states can be alternatively labeled by another string operator
\bea
\notag\hat{O}_L=\sigma^z_{{\bf r}_b}\big(\prod_{\text{R-legs}}\tau^x_{\bf i}\big)\sigma^z_{{\bf r}_e},~~~[\hat{O}_L,H]=0,~~~(\hat{O}_L)^2=1.
\eea
where the open oriented line $L$ starts on one boundary (${\bf r}_b$ is the closet $\vec\sigma$ spin on its r.h.s.) and ends on the other boundary (${\bf r}_e$ on its r.h.s.) in FIG. \ref{fig:torus}. Notice that open line $L$ crosses the closed string $S$ only once, hence $\{\hat{O}_T,\hat{O}_S\}=0$. Both string operators are \emph{even} under Ising symmetry. This means operator $\hat{O}_L$ tunnels between two $\hat{O}_S$ eigenstates in the ground states' manifold:
\bea
\hat{O}_L|O_S=\pm1\rangle=|O_S=\mp1\rangle.\notag
\eea
As we shrink the SET region in FIG. \ref{fig:torus}, the length of line $L$ also decreases and ultimately $\hat{O}_L$ will become a local operator. In this limit the two $\hat{O}_S$ eigenstates start to mix by local interactions and the system picks up a unique ground state as a superposition of $|O_s=\pm1\rangle$ states. As a result we go back to the case of pure $Z_2$-SPT phase, which has no GSD on torus.

Physically how does string operator $\hat{O}_L$ tunnel between the 2 degenerate ground states? There is a clear picture based on effective theory (\ref{boundary term:Z2}). Since the bound states of kinks (from $Z_2$-SPT) and ends of strings (from double semion model) ``condense'' on the boundary, it's number doesn't conserve anymore. physical observables will be screened by the condensate.
The unscreened anyons must have trivial mutual statistics with the condensate, \ie they must commute with the arguments of both cosine terms in (\ref{boundary term:Z2}). It turns out there is only one type of unscreened anyon here:
\bea
\psi=e^{\imth(\phi_s+\phi_{\bar s})}\cdot e^{-\imth\phi_1}\simeq e^{\imth(\phi_s-\phi_{\bar s})}\cdot e^{-\imth\phi_2}\notag
\eea
When such a anyon is created from one boundary, another anyon must be created from the other boundary. The double-semion-model part of this operator (\ie $e^{\imth(\phi_s\pm\phi_{\bar s})}$ part) from both boundary can be brought into the bulk of double-semion-model region and annihilated\cite{Wang2012}, while the $Z_2$-SPT part of operator $\psi$ (\ie $e^{-\imth\phi_{1,2}}$ part) are left on both boundaries. Such a process of of creating (on the boundary) and annihilating (in the SET bulk) anyons is precisely realized by string operator $\hat{O}_L$.\\

\subsection{Quantum spin Hall insulators $\Big(G_s=U(1)\rtimes Z_2^\bst\Big)$}

Now let's turn to a more familiar example of SPT phases of electrons: $\mbz_2$ quantum spin Hall insulator\cite{Kane2005a,Bernevig2006,Konig2007} (QSHI). Its edge theory is (\ref{edge action}) with ${\bf K}=\bpm1&0\\0&-1\epm$. Let's denote the two branches of chiral bosons as $\phi_{R/L}$, and they transform under $U(1)\rtimes Z_2^\bst$ symmetry as
\bea
\notag&e^{\imth\theta\hat{N}_f}\phi_{R/L}e^{-\imth\theta\hat{N}_f}=\phi_{R/L}+\theta,~~~\bpm\phi_R\\ \phi_L\epm\overset{\bst}\rightarrow\bpm-\phi_L\\-\phi_R+\pi\epm.
\eea
Electrons on the edge are $\psi_{R/L}\sim e^{\imth\phi_{R/L}}$. A gap will open up on the edge specrum, by magnetic order $\hat{M}=M\cos(\phi_R-\phi_L+\alpha_M)$ which breaks time reversal $\bst$, or by superconductivity $\hat\Delta=\Delta\cos(\phi_R+\phi_L+\alpha_\Delta)$ which breaks $U(1)$ charge conservation. Can we restore symmetry and obtain a gapped symmetric edge by proliferating defects of order parameters $\hat{M}$ and $\hat\Delta$? The answer is no. For example, domain wall of magnetic order $\hat{D}_\bst=\exp\big[\imth(\phi_R+\phi_L)/2\big]$ is a bosonic object, but it carries unit charge. Therefore proliferating this object will break $U(1)$ charge conservation!

Hence we need to play the same trick and consider the boundary between QSHI and a fractionalized SET phase. A convenient choice is the so-called QSHI$^\ast$ phase\cite{Ruegg2012}, obtained by coupling fermions in QSHI to a dynamical $Z_2$ gauge field\cite{Senthil2000}. As shown in Appendix \ref{app}, the effective edge theory of QSHI$^\ast$ is described by (\ref{edge action}) with ${\bf K}=\bpm0&2\\2&0\epm$. If we label its edge chiral bosons as $\phi_{s/c}$, they transform under $U(1)\rtimes Z_2^\bst$ symmetry as
\bea
\notag&e^{\imth\theta\hat{N}_f}\bpm\phi_s\\ \phi_c\epm e^{-\imth\theta\hat{N}_f}=\bpm\phi_s\\ \phi_c+\theta\epm,\\
&\notag\bpm\phi_s\\ \phi_c\epm\overset{\bst}\longrightarrow\bpm\phi_s\\-\phi_c\epm+\frac{\pi}{2}\bpm1\\1\epm.
\eea
Therefore $e^{\imth\phi_c}$ is a charge-1 boson (``chargeon''), with mutual semion statistics with neutral boson $e^{\imth\phi_s}$ (``spinon'' since $\bst^2=-1$). Clearly $e^{-\imth\phi_c}$ can form a neutral bosonic bound state with domain wall $\hat{D}_\bst$, and its condensation will lead to a gapped symmetric edge between QSHI and QSHI$^\ast$. More precisely the boundary tunneling term is
\bea
\notag&\mathcal{H}_t=T_c\cos(\phi_R+\phi_L-2\phi_c)+T_s\cos(\phi_R-\phi_L-2\phi_s).\\
\label{boundary term:QSHI}
\eea

When we consider the hybrid geometry in FIG. \ref{fig:torus}, there is a 2-fold GSD. Remarkably the two degenerate ground states are labeled by different electron number parity on each edge. Here the closed string operator becomes
\bea
\hat{O}_S=e^{\imth\oint_C\text{d}x~\partial_x\phi_c(x)}=e^{\imth\oint_C\text{d}x~\partial_x\phi_s(x)}.
\eea
where the equality is enforced by boundary condition (\ref{boundary term:QSHI}). Meanwhile the physical meaning of open string operator $\hat{O}_L$ in FIG. \ref{fig:torus} is to create a pair of anyons
\bea
\psi=e^{\imth(\phi_c+\phi_s)}\cdot e^{\imth\phi_R}\simeq e^{\imth(\phi_c-\phi_s)}\cdot e^{\imth\phi_L}
\eea
on each edge, move the $e^{\imth(\phi_c\pm\phi_s)}$ part to the QSHI$^\ast$ bulk and annihilate them. Notice that in this process an extra electron $e^{\imth\phi_{R/L}}$ is created on each boundary, causing the local change of fermion number parity. In supplemental materials we prove that gauging fermion number parity ($Z_2^f$) symmetry in QSHI leads to QSHI$^\ast$, again confirming our conjecture (\ref{conjecture}).

\section{Summary}

In this paper we investigate the question of how to proliferate topological defects (domain walls or kinks) on the symmetry-broken edge of an SPT phase, in order to achieve a gapped symmetric edge. We show that condensing these defects is either forbidden by quantum statistics or breaks symmetry. On the other hand, we can overcome these obstructions by considering a boundary between an SPT phase $\mathcal{S}$ and a fractionalized (SET) phase $\mathcal{F}$. We propose a conjecture (\ref{conjecture}) for how to find such fractionalized phases for a given SPT, which generalizes to all spatial dimensions. Two examples in two spatial dimensions are presented with effective field theory and exactly solvable models. We found that once this hybrid structure is put on a torus as in FIG. \ref{fig:torus}, there will be ground state degeneracy accompanying the two gapped symmetric edges between $\mathcal{S}$ and $\mathcal{F}$.

\emph{Note added}~~~Upon completion of this work. we became aware of \Ref{Wang2013}, where a different geometry containing both SPT and SET phases are considered. In their situation open string order parameter $\hat{O}_L$ only contains operators from SET side on one boundary, therefore it breaks symmetry in a nonlocal way.

\acknowledgements

This work is supported by DOE Office
of Basic Energy Sciences, Materials Sciences Division of the U.S. DOE under contract No. DE-AC02-05CH11231 (YML,DHL) and in part by the National Science Foundation under Grant No. PHYS-1066293(YML). We thank Taylor L. Hughes and Ashvin Vishwanath for helpful conversations. YML thanks the hospitality of the Aspen Center for Physics where he presented this work in 2013 summer program ``Disorder, Dynamics, Frustration and Topology in Quantum Condensed Matter", as well as Institute for Advance Study in Tsinghua University for hospitality where part of this work is finished.

\appendix

\section{Effective field theory of QSHI$^\ast$}\label{app}

QSHI$^\ast$ is obtained by gauging $Z_2^f$ (fermion parity) symmetry (generated by $\bsg=(-1)^{\hat{N}_f}$) in a fermionic QSHI in 2+1-D. This means we need to couple a dynamical $Z_2$ gauge field to fermions in QSHI. Following discussions in the previous section, in this case we have
\bea\label{kmat:qshi}
{\bf K}=\bpm1&0\\0&-1\epm,~~~\vec{\delta\phi^\bsg}=\bpm\pi\\ \pi\epm\Longrightarrow{\bf l}^\bsg=\frac12\bpm1\\-1\epm.
\eea
Hence after gauging $Z_2^f$ symmetry, we obtained a fractionalized SET phase $\mathcal{F}$ with
\bea
&{\bf K}^\bsg={\bf M}^{-1}{\bf K}\big({\bf M}^{-1}\big)^T=\bpm4&2\\2&0\epm,\\
&\notag{\bf M}=\bpm1/2&0\\-1/2&1\epm.
\eea
We can always make a ${\bf X}\in GL(2,\mbz)$ rotation\cite{Lu2012a} to the ${\bf K}$ matrices so that
\bea\label{kmat:qshi star}
&{\bf K}^\bsg\simeq{\bf X}^T{\bf K}^\bsg{\bf X}=\bpm0&2\\2&0\epm,\\
&\notag{\bf X}=\bpm1&0\\-1&1\epm.
\eea
Let's label the chiral boson fields in QSHI (\ref{kmat:qshi}) as $\phi_{R/L}$, and those in QSHI$^\ast$ (\ref{kmat:qshi star}) as $\phi_{c/s}$. They have the following correspondence:
\bea
\bpm\phi_s\\ \phi_c\epm\leftrightarrow{\bf X}^{-1}{\bf M}^T\bpm\phi_R\\ \phi_L\epm.
\eea
Hence one can easily figure out the symmetry transformations of quasiparticles in QSHI$^ast$
\bea
&e^{\imth\theta\hat{N}_f}\bpm\phi_s\\ \phi_c\epm e^{-\imth\theta\hat{N}_f}=\bpm\phi_s\\ \phi_c+\theta\epm,\\
&\notag\bpm\phi_s\\ \phi_c\epm\overset{\bst}\longrightarrow\bpm\phi_s\\-\phi_c\epm+\frac{\pi}{2}\bpm1\\1\epm.
\eea

%
%



\begin{thebibliography}{31}
\expandafter\ifx\csname natexlab\endcsname\relax\def\natexlab#1{#1}\fi
\expandafter\ifx\csname bibnamefont\endcsname\relax
  \def\bibnamefont#1{#1}\fi
\expandafter\ifx\csname bibfnamefont\endcsname\relax
  \def\bibfnamefont#1{#1}\fi
\expandafter\ifx\csname citenamefont\endcsname\relax
  \def\citenamefont#1{#1}\fi
\expandafter\ifx\csname url\endcsname\relax
  \def\url#1{\texttt{#1}}\fi
\expandafter\ifx\csname urlprefix\endcsname\relax\def\urlprefix{URL }\fi
\providecommand{\bibinfo}[2]{#2}
\providecommand{\eprint}[2][]{\url{#2}}

\bibitem[{\citenamefont{Hasan and Kane}(2010)}]{Hasan2010}
\bibinfo{author}{\bibfnamefont{M.~Z.} \bibnamefont{Hasan}} \bibnamefont{and}
  \bibinfo{author}{\bibfnamefont{C.~L.} \bibnamefont{Kane}},
  \bibinfo{journal}{Rev. Mod. Phys.} \textbf{\bibinfo{volume}{82}},
  \bibinfo{pages}{3045} (\bibinfo{year}{2010}).

\bibitem[{\citenamefont{Moore}(2010)}]{Moore2010}
\bibinfo{author}{\bibfnamefont{J.~E.} \bibnamefont{Moore}},
  \bibinfo{journal}{Nature} \textbf{\bibinfo{volume}{464}},
  \bibinfo{pages}{194} (\bibinfo{year}{2010}), ISSN \bibinfo{issn}{0028-0836}.

\bibitem[{\citenamefont{Qi and Zhang}(2011)}]{Qi2011}
\bibinfo{author}{\bibfnamefont{X.-L.} \bibnamefont{Qi}} \bibnamefont{and}
  \bibinfo{author}{\bibfnamefont{S.-C.} \bibnamefont{Zhang}},
  \bibinfo{journal}{Rev. Mod. Phys.} \textbf{\bibinfo{volume}{83}},
  \bibinfo{pages}{1057} (\bibinfo{year}{2011}).

\bibitem[{\citenamefont{Chen et~al.}(2013)\citenamefont{Chen, Gu, Liu, and
  Wen}}]{Chen2013a}
\bibinfo{author}{\bibfnamefont{X.}~\bibnamefont{Chen}},
  \bibinfo{author}{\bibfnamefont{Z.-C.} \bibnamefont{Gu}},
  \bibinfo{author}{\bibfnamefont{Z.-X.} \bibnamefont{Liu}}, \bibnamefont{and}
  \bibinfo{author}{\bibfnamefont{X.-G.} \bibnamefont{Wen}},
  \bibinfo{journal}{Phys. Rev. B} \textbf{\bibinfo{volume}{87}},
  \bibinfo{pages}{155114} (\bibinfo{year}{2013}).

\bibitem[{\citenamefont{Mermin}(1979)}]{Mermin1979}
\bibinfo{author}{\bibfnamefont{N.~D.} \bibnamefont{Mermin}},
  \bibinfo{journal}{Rev. Mod. Phys.} \textbf{\bibinfo{volume}{51}},
  \bibinfo{pages}{591} (\bibinfo{year}{1979}).

\bibitem[{\citenamefont{Wilczek}(1990)}]{Wilczek1990B}
\bibinfo{author}{\bibfnamefont{F.}~\bibnamefont{Wilczek}},
  \emph{\bibinfo{title}{Fractional Statistics and Anyon Superconductivity}}
  (\bibinfo{publisher}{World Scientific Pub Co Inc}, \bibinfo{year}{1990}).

\bibitem[{\citenamefont{Lu and Vishwanath}(2012)}]{Lu2012a}
\bibinfo{author}{\bibfnamefont{Y.-M.} \bibnamefont{Lu}} \bibnamefont{and}
  \bibinfo{author}{\bibfnamefont{A.}~\bibnamefont{Vishwanath}},
  \bibinfo{journal}{Phys. Rev. B} \textbf{\bibinfo{volume}{86}},
  \bibinfo{pages}{125119} (\bibinfo{year}{2012}).

\bibitem[{\citenamefont{Read}(1990)}]{Read1990}
\bibinfo{author}{\bibfnamefont{N.}~\bibnamefont{Read}}, \bibinfo{journal}{Phys.
  Rev. Lett.} \textbf{\bibinfo{volume}{65}}, \bibinfo{pages}{1502}
  (\bibinfo{year}{1990}).

\bibitem[{\citenamefont{Wen and Zee}(1992)}]{Wen1992}
\bibinfo{author}{\bibfnamefont{X.~G.} \bibnamefont{Wen}} \bibnamefont{and}
  \bibinfo{author}{\bibfnamefont{A.}~\bibnamefont{Zee}},
  \bibinfo{journal}{Phys. Rev. B} \textbf{\bibinfo{volume}{46}},
  \bibinfo{pages}{2290} (\bibinfo{year}{1992}).

\bibitem[{\citenamefont{Frohlich and Zee}(1991)}]{Frohlich1991}
\bibinfo{author}{\bibfnamefont{J.}~\bibnamefont{Frohlich}} \bibnamefont{and}
  \bibinfo{author}{\bibfnamefont{A.}~\bibnamefont{Zee}},
  \bibinfo{journal}{Nuclear Physics B} \textbf{\bibinfo{volume}{364}},
  \bibinfo{pages}{517} (\bibinfo{year}{1991}), ISSN \bibinfo{issn}{0550-3213}.

\bibitem[{\citenamefont{Wen}(1995)}]{Wen1995}
\bibinfo{author}{\bibfnamefont{X.-G.} \bibnamefont{Wen}},
  \bibinfo{journal}{Advances in Physics} \textbf{\bibinfo{volume}{44}},
  \bibinfo{pages}{405} (\bibinfo{year}{1995}), ISSN \bibinfo{issn}{0001-8732}.

\bibitem[{\citenamefont{Levin and Stern}(2012)}]{Levin2012a}
\bibinfo{author}{\bibfnamefont{M.}~\bibnamefont{Levin}} \bibnamefont{and}
  \bibinfo{author}{\bibfnamefont{A.}~\bibnamefont{Stern}},
  \bibinfo{journal}{Phys. Rev. B} \textbf{\bibinfo{volume}{86}},
  \bibinfo{pages}{115131} (\bibinfo{year}{2012}).

\bibitem[{\citenamefont{Levin and Gu}(2012)}]{Levin2012}
\bibinfo{author}{\bibfnamefont{M.}~\bibnamefont{Levin}} \bibnamefont{and}
  \bibinfo{author}{\bibfnamefont{Z.-C.} \bibnamefont{Gu}},
  \bibinfo{journal}{Phys. Rev. B} \textbf{\bibinfo{volume}{86}},
  \bibinfo{pages}{115109} (\bibinfo{year}{2012}).

\bibitem[{\citenamefont{Chen et~al.}(2014)\citenamefont{Chen, Lu, and
  Vishwanath}}]{Chen2014}
\bibinfo{author}{\bibfnamefont{X.}~\bibnamefont{Chen}},
  \bibinfo{author}{\bibfnamefont{Y.-M.} \bibnamefont{Lu}}, \bibnamefont{and}
  \bibinfo{author}{\bibfnamefont{A.}~\bibnamefont{Vishwanath}},
  \bibinfo{journal}{Nat Commun} \textbf{\bibinfo{volume}{5}},
  (\bibinfo{year}{2014}).

\bibitem[{\citenamefont{Vishwanath and Senthil}(2013)}]{Vishwanath2013}
\bibinfo{author}{\bibfnamefont{A.}~\bibnamefont{Vishwanath}} \bibnamefont{and}
  \bibinfo{author}{\bibfnamefont{T.}~\bibnamefont{Senthil}},
  \bibinfo{journal}{Phys. Rev. X} \textbf{\bibinfo{volume}{3}},
  \bibinfo{pages}{011016} (\bibinfo{year}{2013}).

\bibitem[{\citenamefont{Wen}(2004)}]{Wen2004B}
\bibinfo{author}{\bibfnamefont{X.-G.} \bibnamefont{Wen}},
  \emph{\bibinfo{title}{Quantum Field Theory Of Many-body Systems: From The
  Origin Of Sound To An Origin Of Light And Electrons}}
  (\bibinfo{publisher}{Oxford University Press, New York},
  \bibinfo{year}{2004}).

\bibitem[{\citenamefont{Essin and Hermele}(2013)}]{Essin2013}
\bibinfo{author}{\bibfnamefont{A.~M.} \bibnamefont{Essin}} \bibnamefont{and}
  \bibinfo{author}{\bibfnamefont{M.}~\bibnamefont{Hermele}},
  \bibinfo{journal}{Phys. Rev. B} \textbf{\bibinfo{volume}{87}},
  \bibinfo{pages}{104406} (\bibinfo{year}{2013}).

\bibitem[{\citenamefont{Mesaros and Ran}(2013)}]{Mesaros2013}
\bibinfo{author}{\bibfnamefont{A.}~\bibnamefont{Mesaros}} \bibnamefont{and}
  \bibinfo{author}{\bibfnamefont{Y.}~\bibnamefont{Ran}},
  \bibinfo{journal}{Phys. Rev. B} \textbf{\bibinfo{volume}{87}},
  \bibinfo{pages}{155115} (\bibinfo{year}{2013}).

\bibitem[{\citenamefont{Hung and Wen}(2013)}]{Hung2013}
\bibinfo{author}{\bibfnamefont{L.-Y.} \bibnamefont{Hung}} \bibnamefont{and}
  \bibinfo{author}{\bibfnamefont{X.-G.} \bibnamefont{Wen}},
  \bibinfo{journal}{Phys. Rev. B} \textbf{\bibinfo{volume}{87}},
  \bibinfo{pages}{165107} (\bibinfo{year}{2013}).

\bibitem[{\citenamefont{Lu and Vishwanath}(2013)}]{Lu2013}
\bibinfo{author}{\bibfnamefont{Y.-M.} \bibnamefont{Lu}} \bibnamefont{and}
  \bibinfo{author}{\bibfnamefont{A.}~\bibnamefont{Vishwanath}},
  \bibinfo{journal}{ArXiv e-prints 1302.2634}  (\bibinfo{year}{2013}),
  \eprint{1302.2634}.

\bibitem[{\citenamefont{Hung and Wan}(2013)}]{Hung2013a}
\bibinfo{author}{\bibfnamefont{L.-Y.} \bibnamefont{Hung}} \bibnamefont{and}
  \bibinfo{author}{\bibfnamefont{Y.}~\bibnamefont{Wan}},
  \bibinfo{journal}{Phys. Rev. B} \textbf{\bibinfo{volume}{87}},
  \bibinfo{pages}{195103} (\bibinfo{year}{2013}).

\bibitem[{\citenamefont{Hung and {Wen}}(2012)}]{Hung2012a}
\bibinfo{author}{\bibfnamefont{L.-Y.} \bibnamefont{Hung}} \bibnamefont{and}
  \bibinfo{author}{\bibfnamefont{X.-G.} \bibnamefont{{Wen}}},
  \bibinfo{journal}{ArXiv e-prints 1211.2767}  (\bibinfo{year}{2012}),
  \eprint{1211.2767}.

\bibitem[{\citenamefont{Levin and Wen}(2005)}]{Levin2005}
\bibinfo{author}{\bibfnamefont{M.~A.} \bibnamefont{Levin}} \bibnamefont{and}
  \bibinfo{author}{\bibfnamefont{X.-G.} \bibnamefont{Wen}},
  \bibinfo{journal}{Phys. Rev. B} \textbf{\bibinfo{volume}{71}},
  \bibinfo{pages}{045110} (\bibinfo{year}{2005}).

\bibitem[{\citenamefont{Kitaev}(2003)}]{Kitaev2003}
\bibinfo{author}{\bibfnamefont{A.~Y.} \bibnamefont{Kitaev}},
  \bibinfo{journal}{Annals of Physics} \textbf{\bibinfo{volume}{303}},
  \bibinfo{pages}{2} (\bibinfo{year}{2003}), ISSN \bibinfo{issn}{0003-4916}.

\bibitem[{\citenamefont{{Wang} and {Wen}}(2012)}]{Wang2012}
\bibinfo{author}{\bibfnamefont{J.}~\bibnamefont{{Wang}}} \bibnamefont{and}
  \bibinfo{author}{\bibfnamefont{X.-G.} \bibnamefont{{Wen}}},
  \bibinfo{journal}{ArXiv e-prints}  (\bibinfo{year}{2012}),
  \eprint{1212.4863}.

\bibitem[{\citenamefont{Kane and Mele}(2005)}]{Kane2005a}
\bibinfo{author}{\bibfnamefont{C.~L.} \bibnamefont{Kane}} \bibnamefont{and}
  \bibinfo{author}{\bibfnamefont{E.~J.} \bibnamefont{Mele}},
  \bibinfo{journal}{Phys. Rev. Lett.} \textbf{\bibinfo{volume}{95}},
  \bibinfo{pages}{146802} (\bibinfo{year}{2005}).

\bibitem[{\citenamefont{Bernevig and Zhang}(2006)}]{Bernevig2006}
\bibinfo{author}{\bibfnamefont{B.~A.} \bibnamefont{Bernevig}} \bibnamefont{and}
  \bibinfo{author}{\bibfnamefont{S.-C.} \bibnamefont{Zhang}},
  \bibinfo{journal}{Phys. Rev. Lett.} \textbf{\bibinfo{volume}{96}},
  \bibinfo{pages}{106802} (\bibinfo{year}{2006}).

\bibitem[{\citenamefont{Konig et~al.}(2007)\citenamefont{Konig, Wiedmann,
  Brune, Roth, Buhmann, Molenkamp, Qi, and Zhang}}]{Konig2007}
\bibinfo{author}{\bibfnamefont{M.}~\bibnamefont{Konig}},
  \bibinfo{author}{\bibfnamefont{S.}~\bibnamefont{Wiedmann}},
  \bibinfo{author}{\bibfnamefont{C.}~\bibnamefont{Brune}},
  \bibinfo{author}{\bibfnamefont{A.}~\bibnamefont{Roth}},
  \bibinfo{author}{\bibfnamefont{H.}~\bibnamefont{Buhmann}},
  \bibinfo{author}{\bibfnamefont{L.~W.} \bibnamefont{Molenkamp}},
  \bibinfo{author}{\bibfnamefont{X.-L.} \bibnamefont{Qi}}, \bibnamefont{and}
  \bibinfo{author}{\bibfnamefont{S.-C.} \bibnamefont{Zhang}},
  \bibinfo{journal}{Science} \textbf{\bibinfo{volume}{318}},
  \bibinfo{pages}{766} (\bibinfo{year}{2007}),
  \eprint{http://www.sciencemag.org/content/318/5851/766.full.pdf}.

\bibitem[{\citenamefont{Ruegg and Fiete}(2012)}]{Ruegg2012}
\bibinfo{author}{\bibfnamefont{A.}~\bibnamefont{Ruegg}} \bibnamefont{and}
  \bibinfo{author}{\bibfnamefont{G.~A.} \bibnamefont{Fiete}},
  \bibinfo{journal}{Phys. Rev. Lett.} \textbf{\bibinfo{volume}{108}},
  \bibinfo{pages}{046401} (\bibinfo{year}{2012}).

\bibitem[{\citenamefont{Senthil and Fisher}(2000)}]{Senthil2000}
\bibinfo{author}{\bibfnamefont{T.}~\bibnamefont{Senthil}} \bibnamefont{and}
  \bibinfo{author}{\bibfnamefont{M.~P.~A.} \bibnamefont{Fisher}},
  \bibinfo{journal}{Phys. Rev. B} \textbf{\bibinfo{volume}{62}},
  \bibinfo{pages}{7850} (\bibinfo{year}{2000}).

\bibitem[{\citenamefont{Wang and Levin}(2013)}]{Wang2013}
\bibinfo{author}{\bibfnamefont{C.}~\bibnamefont{Wang}} \bibnamefont{and}
  \bibinfo{author}{\bibfnamefont{M.}~\bibnamefont{Levin}},
  \bibinfo{journal}{Phys. Rev. B} \textbf{\bibinfo{volume}{88}},
  \bibinfo{pages}{245136} (\bibinfo{year}{2013}).

\end{thebibliography}

\end{document}